# SCVCNet: Sliding cross-vector convolution network for cross-task and inter-individual-set EEG-based cognitive workload recognition

Qi Wang, Li Chen, Zhiyuan Zhan, Jianhua Zhang, and Zhong Yin,

*Abstract*—This paper presents a generic approach for applying the cognitive workload recognizer by exploiting common electroencephalogram (EEG) patterns across different human-machine tasks and individual sets. We propose a neural network called SCVCNet, which eliminates task- and individual-set-related interferences in EEGs by analyzing finer-grained frequency structures in the power spectral densities. The SCVCNet utilizes a sliding cross-vector convolution (SCVC) operation, where paired input layers representing the theta and alpha power are employed. By extracting the weights from a kernel matrix's central row and column, we compute the weighted sum of the two vectors around a specified scalp location. Next, we introduce an inter-frequency-point feature integration module to fuse the SCVC feature maps. Finally, we combined the two modules with the output-channel pooling and classification layers to construct the model. To train the SCVCNet, we employ the regularized least-square method with ridge regression and the extreme learning machine theory. We validate its performance using three databases, each consisting of distinct tasks performed by independent participant groups. The average accuracy (0.6813 and 0.6229) and F1 score (0.6743 and 0.6076) achieved in two different validation paradigms show partially higher performance than the previous works. All features and algorithms are available on website: https://github.com/7ohnKeats/SCVCNet.

## I. INTRODUCTION

Previous studies have defined cognitive workload as the dynamic occupation of cognitive capacity when an individual achieves a certain performance in a given task, influenced by various scenario factors [1]. In the context of human-machine collaboration, it becomes crucial for human operators to effectively allocate their cognitive capacity to task-related information in order to sustain overall performance. In safety-critical tasks characterized by stringent performance specifications, such as aircraft driving [2], medical care systems [3], and monitoring of chemical plants [4], cognitive capacity overload can occur before performance degradation, potentially leading to catastrophic accidents [2]-

[4]. Therefore, measuring cognitive workload can play a pivotal role in predicting the risk of human-factor failures, thereby ensuring the safety of human operators and the task environment. Specifically, the electroencephalogram (EEG) possesses qualities of objectivity and repeatability, facilitating the continuous evaluation of responses originating from the central nervous system and cognitive functionality [5]-[8], thus serving as a valuable external indicator for monitoring fluctuations in cognitive workload.

Passive brain-computer interfaces (BCIs) and machine learning approaches have been employed to acquire and decode EEG signals for evaluating cognitive workload in previous works [5]-[8]. In this context, the common approach for EEG decoding involves decomposing the signals into delta (1–4 Hz), theta (4–8 Hz), alpha (8–12 Hz), beta (12–30 Hz), and gamma (>30 Hz) frequency bands, allowing individual examination of the power spectral density (PSD) in each band [9]. Among the different PSD bands, previous studies have demonstrated that theta and alpha rhythms are closely associated with cognitive and memory processes [9]. This correlation is primarily observed in two aspects: 1) a low level of cognitive workload is indicated by a tonic increase in alpha power accompanied by a decrease in theta power; 2) a higher level of cognitive workload is reflected by a substantial decrease in alpha power accompanied by an increase in theta power [10].

Reported studies have primarily evaluated the performance of cognitive workload recognizers in the context of a single human-machine task or a fixed set of individuals [6]. However, in real-world scenarios, workload recognition is often applied to diverse cognitive tasks involving different groups of operators. The variation in task type and individual groups can have an impact on the distribution of cortical spectral activity [11]. When individuals interact with different task types, there are variations in the peak values of PSD due to differences in the degree of mental resource occupancy [12]. Moreover, the centers of PSD distributions on the scalp can differ across multiple individual groups. Two different individual groups may exhibit a decrease in alpha power with increasing task difficulty either in the parietal region or in both parietal and frontal regions [13]. In the existing literature, less attention has

[1] This work is sponsored by the National Natural Science Foundation of China under Grant No. 61703277 and the Shanghai Sailing Program under Grant No. 17YF1427000.

Q. Wang, L. Chen, and Z. Zhan are with the Department of Control Science and Engineering the School of Optical-Electrical and Computer Engineering, University of Shanghai for Science and Technology, Shanghai, 200093, P. R. China (e-mail: 212240441@st.usst.edu.cn; 223330773@st.usst.edu.cn; 223330818@st.usst.edu.cn).

J. Zhang is with the OsloMet Artificial Intelligence Lab, Department of Computer Science, Oslo Metropolitan University, Oslo, N-0130, Norway (e-mail: jianhuaz@oslomet.no).

Corresponding author, Z. Yin is with the Engineering Research Center of Optical Instrument and System, Ministry of Education, Shanghai Key Lab of Modern Optical System, University of Shanghai for Science and Technology, Shanghai, 200093, P. R. China (e-mail: yinzhong@usst.edu.cn).

been given to simultaneously addressing the challenges of cross-task and inter-individual-set interferences when recognizing cognitive workload levels.

To tackle this issue, the conventional approach involves the utilization of transfer learning combined with domain adaptation methodologies. Nevertheless, it necessitates the utilization of EEG samples to capture distributional characteristics from the testing dataset, a task that may prove challenging in practical situations. In scenarios where only training data are accessible, machine learning techniques must possess the capacity to represent PSD features in a manner that adeptly retains workload-related information while eliminating task- and individual-specific components. Since the spatial pattern of the EEG is constrained by the limited number of electrodes, and the temporal pattern can be distorted by various sources of artifacts, one possible solution is to examine the finer-grained frequency structures within the PSD features. However, classical machine learning models such as support vector machine (SVM), random forest (RF), multilayer perceptron (MLP), convolutional neural network (ConvNet), and long-short term memory network (LSTM) are not explicitly designed for this purpose [14]. A possible solution is to build a feedforward mapping as the filter that can identify stable workload-related information by leveraging the finer-grained PSD calculated across consecutive frequency point pairs in the workload-related bands.

Based on above motivation, the contributions of the present study can be summarized as follows:

(1) We assess the performance of cognitive workload recognizers under the challenging conditions of cross-task and inter-individual-set interferences. To accomplish this, we employ three publicly available databases: the N-back database (collected in our previous work) [15], the Simultaneous Task EEG Workload database [16], and the Electroencephalograms during Mental Arithmetic Task database [17]. These databases consist of the N-back, simultaneous capacity test, and mental arithmetic tasks performed by three independent groups of participants. In total, we thoroughly investigate the performance of workload recognizers using two training and testing paradigms, covering six combinations of the databases.

(2) We propose a novel neural network architecture called SCVCNet, designed to address the cross-task and inter-individual-set interferences in EEG signals. The key component of SCVCNet is the sliding cross-vector convolution (SCVC) operation. It utilizes paired input layers representing finer-grained theta and alpha power vectors. By extracting the weights from a kernel matrix's central row and column, we compute the weighted sum of the two vectors around a specified scalp location. Next, we introduce an inter-frequency-point feature integration module to fuse the SCVC feature maps among all frequency sample pairs. To improve training speed, we further adopt the regularized least-square method using ridge regression and the theory of extreme learning machines (ELM) instead of the adaptive momentum (ADAM) optimization algorithm.

The remaining sections of this paper are organized as follows: Section II reviews previous works. Section III describes the databases and EEG features. Section IV elaborates on the details of the SCVCNet architecture. Section V presents the evaluation of the cross-task and inter-individual-set workload recognizers and provides corresponding results. Section VI includes useful discussions. Section VII summarizes the findings and limitations of the present study.

## II. RELATED WORKS

Previous studies have widely investigated EEG-based cross-task cognitive recognition, utilizing an identical set of participants. Baldwin et al. [18] recorded EEGs derived from three distinct cognitive tasks, namely, the reading span, visuospatial N-back, and Sternberg memory scanning task, conducted by fifteen subjects. The results derived by a single-hidden layer feedforward neural network demonstrated an average classification accuracy of 44.8% for binary workload levels across the three tasks. Gómez et al. [19] assembled 19 subjects and collected EEG recordings during tasks involving the N-back and Stroop [20] paradigms. They accomplished a cross-task recognition accuracy of 55.64% for binary workload levels, utilizing a RF classifier. Guan et al. [21] recorded the EEG data from 16 participants engaged in the N-back task under four distinct types of stimulation, encompassing verbal, object, spatial verbal, and spatial object conditions. Employing dynamic functional connectivity metrics in tandem with an SVM-based classification strategy, they achieved a three-class accuracy of 80.3%.

Subsequently, the investigators augmented the size of the participant pool. Dimitrakopoulos et al. [22] amassed EEG signals from 28 participants who undertook both N-back and arithmetic tasks. The analytical approach commenced with the extraction of cortical functional connections, followed by the application of sequential forward feature selection. They employed an SVM for the classification of these selected features, ultimately achieving an accuracy of 87% for binary workload levels. Kakkos et al. [12] recorded the EEGs from 40 volunteers who actively engaged in the N-back and mental arithmetic tasks. Based on the PSDs, functional connectivity, and an SVM for classification, their framework culminated in a binary accuracy rate of 94%. In a parallel study, Gupta et al. [23] curated EEG data from 44 subjects who participated in two visual tasks, namely, the recognition of diverse shapes and colors. Employing an intricate deep recurrent neural network architecture, they achieved an accuracy of 92.8%.

Recently, deep ConvNets, LSTMs, and Transformers have been used as backbones of the workload recognizer. Zhang et al. [24] procured EEG data from 20 male participants engaged in the N-back and arithmetic tasks. They introduced a strategy by proposing a pseudo-3D-ConvNet integrated with the venerable recurrent neural network to streamline the parameter set and reached an accuracy of 88.9%. Zhou et al. [25] undertook an investigation by collecting EEGs from 45 subjects who participated in letter sequence memorization and mental arithmetic tasks. In their study, they explored the efficacy of four domain adaptation techniques, namely transfer component analysis, joint distribution adaptation, balanced domain adaptation, and transfer joint matching. Their findings revealed that the application of transfer learning methodologies led to a notable improvement in accuracy, with gains ranging from 3% to 8%. In our prior study [15], we assembled EEG data from 20 participants actively engaged in a dual N-back task. We devised a dynamic residual network incorporated with an attention mechanism, aptly named the DRNANet, to facilitate binary cross-task cognitive workload classification. This innovative architecture was employed to classify both PSD and statistical features, ultimately achieving a classification accuracy of 60.55%.

## III. MATERIALS

### A. Data Acquisition

In our previous study [15], we conducted an experiment involving 20 participants to investigate their EEG signals while performing a dual N-back task. Prior to data acquisition, we provided the participants with training on the task's manual operations and familiarized them with the experimental procedure. Each participant initially underwent a pre-resting phase, which lasted for 180 seconds. Following this, the participants engaged in the dual N-back task, experiencing low, moderate, and high complexities, each lasting 180 seconds. Finally, a post-resting phase of 180 seconds was included. During the task phases, the participants were instructed to compare the position, colors, and shapes of visual images displayed on the screen at the current and previous time steps, determining whether they were the same or different. Simultaneously, they were also required to assess the consistency of the accompanying audio information. Throughout all five phases, the participants' EEG signals were recorded and used to construct the N-back database.

In the study conducted by Lim et al. [16], a total of 50 male participants were recruited to complete a Simultaneous Capacity (SIMKAP) test task using the Vienna Test System [26]. This task required participants to compare two panels and cross out the identical items. Additionally, participants were later asked to answer several auditory questions. Each participant began by sitting comfortably for a resting phase lasting 180 seconds. Following this, they performed the SIMKAP task phase for another 180 seconds. Throughout all phases, the EEG signals were recorded. However, the EEG data of two participants were excluded due to incomplete task performance. Furthermore, the first and last 15 seconds of EEG recordings from each phase were removed and deemed unavailable. Consequently, only the remaining 150-second EEG recordings from each phase, combined with the resting phase data, were used to construct the Simultaneous Task EEG Workload (STEW) database.

In the study conducted by Zyma et al. [17], a total of 36 participants were engaged in a mental arithmetic task involving continuous subtraction. Prior to the experiment, participants underwent a 180-second training phase to familiarize themselves with the subtraction operations. Following the training phase, participants had a comfortable resting phase lasting 180 seconds. Subsequently, they performed the subtraction task phase for a duration of 240 seconds. EEG signals were recorded during both the resting and task phases. It is worth noting that in the subtraction task phase, the EEG data from the first 60 seconds were available in [17]. Finally, the EEG data collected from all the participants during the resting phase and the selected task phase were utilized to create the mental arithmetic (EEGMAT) database.

TABLE I
ASSIGNMENT OF THE TARGET CLASSES.

| Databases | N-back | STEW | EEGMAT |
|---|---|---|---|
| Participants | 20 | 48 | 36 |
| EEG channels | 14 | 14 | 19 |
| Sampling rate | 128 Hz | 128 Hz | 500 Hz |
| Type of the task | Dual N-back | SIMKAP | Mental arithmetic |
| Unloaded class | Post-resting phase | Resting phase | Resting phase |
| Loaded class | Moderate phase | Task phase | Task phase |
| UL duration | 180 s | 150 s | 180 s |
| Selection of UL | 60 s ~ 120 s | 45 s ~ 105 s | 60 s ~ 120 s |
| LD duration | 180 s | 150 s | 60 s |
| Selection of LD | 60 s ~ 120 s | 45 s ~ 105 s | 0 s ~ 60 s |

Note: UL and LD denote the unloaded and loaded classes of the cognitive workload.

### B. Assignment of the Target Classes of the Cognitive Workload

We conducted a cross-task cognitive workload recognition study with the objective of distinguishing between unloaded and loaded classes, utilizing data from the N-back, STEW, and EEGMAT databases. The allocation of target classes for binary workload levels in the EEG epochs is detailed in Table I. In the case of the N-back database, we designated the post-resting and moderate complexity phases as representing the unloaded and loaded classes, respectively. This decision was informed by participant feedback, which revealed that individuals often experienced difficulty achieving complete relaxation during the pre-resting phase due to heightened nervousness. Additionally, the low complexity phase was deemed insufficient in eliciting substantial task demands, as indicated by participants' subjective ratings. Conversely, the high-complexity phase exhibited noticeable performance deterioration, potentially imposing additional psychological burdens such as negative emotions on participants. Consequently, we retained the post-resting and moderate complexity phases for the N-back database. In contrast, for the STEW and EEGMAT databases, only a single resting phase and a single task phase were available. These two phases were directly assigned as representing the unloaded and loaded classes, respectively.

We opted for EEG signal segments lasting 60 seconds, specifically targeting the midpoint of each phase. The EEGMAT database presented a unique scenario, offering only a 60-second EEG segment from the task phase. To maintain a sufficient quantity of EEG segments and ensure class balance for machine learning model training, we made the choice to utilize the entire duration of signals from the task phase. Additionally, our decision to focus on the middle section of the phases stemmed from considerations regarding participant engagement dynamics. At the beginning of the task phase, there may be low engagement, while towards the end, participants could experience heightened mental fatigue.

### C. Channel Selection

In both the N-back and STEW databases, EEG recordings were acquired using the Emotiv EPOC device, comprising 14 channels, with a sampling frequency of 128 Hz. Conversely, the EEGMAT database employed the Neurocom EEG device, equipped with 19 channels, and operated at a sampling frequency of 500 Hz. The channel configurations adhered to the standards of the International 10-20 System, as illustrated in Fig.

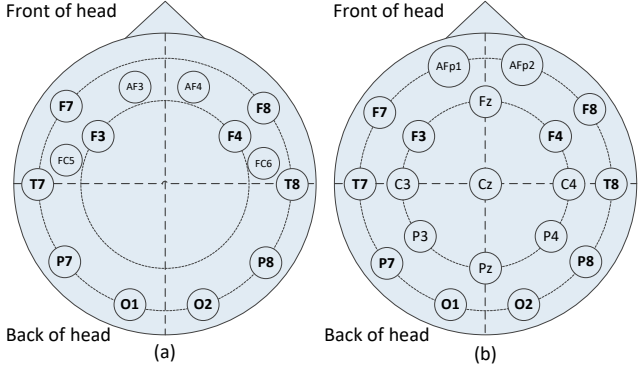

**Fig. 1.** Channel locations of the three databases. The EEG channels of the N-back and STEW are shown in (a) and that of the EEGMAT is shown in (b). The common channel locations are marked in bold.

1. It is worth noting that ten channels (F3, F7, T7, P7, O1, O2, P8, T8, F8, and F4) were common across all three databases. We then focused our analysis on the EEG data from these common channels. This approach was deemed appropriate given the sensitivity of cognitive workload variation to frontal theta power (F3, F7, F8, and F4) and parietal alpha power (P7 and P8) [9]. Additionally, channels C3, Cz, and C4, which are associated with motor responses, were excluded from analysis due to the potential interference arising from manual task execution.

### D. EEG Preprocessing and Feature Extraction

Subsequently, uniform preprocessing procedures were applied to the 10-channel EEG data across all three databases. Initially, we addressed outliers in the raw signal of each channel by assessing the absolute value of the first-order signal difference based on a predefined threshold. Any identified aberrant values were replaced with the nearest preceding normal values. Following this step, we implemented a finite impulse response (FIR) filter, employing a passband ranging from 3.5 Hz to 31 Hz and applying the Hamming window for signal filtering. The specified low and high cutoff frequencies were selected to mitigate ocular and muscular artifacts. In order to maintain a consistent sampling frequency with the other two databases, we downsampled the EEGMAT signals from 500 Hz to 128 Hz.

Following the preprocessing phase, the 60-second EEG signals for each phase were partitioned into segments utilizing a 20-second rectangular window, incorporating a stride of 10 seconds. Consequently, every two consecutive segments shared a 10-second overlap. Within each of these segments, we extracted 32 PSD values spanning the frequency range of 4 Hz to 12 Hz. The selection of this frequency band was underpinned by the proximity of theta (4 Hz to 7 Hz) and alpha power (8 Hz to 12 Hz) to variations in cognitive workload levels. The frequency resolution of the PSD features amounted to 4 points per Hz. The computation of the PSDs was carried out employing the Welch method [27]. This involved Welch segments of length 128 points, an overlap of 64 points between each segment, and a Hamming window of 512 points for the Fast Fourier Transform.

## IV. SLIDING CROSS-VECTOR CONVOLUTION NETWORK

The overall feedforward architecture of the SCVCNet for mapping the PSDs to estimated cognitive workload levels is shown in Fig. 2. The following subsections elaborate the details of each component and its training algorithm.

### A. Paired Input Layers

There are two parallel input layers of the SCVCNet. For the input layer #1, the EEG features are organized as a matrix $\mathbf{X} \in \mathcal{R}^{p \times m}$, where $p$ and $m$ denote the numbers of the PSD features in the theta band (4~7 Hz) and the selected EEG channels, respectively. For the input layer #2, we obtain a feature matrix $\mathbf{Y} \in \mathcal{R}^{q \times n}$, where $q$ and $n$ denote the numbers of the PSD features in the alpha band (8~12 Hz) and the selected channels, respectively. In the present study, we have $p = q = 16$ and $m = n = 10$. Namely, we transferred sixteen PSD values and ten shared EEG channels for each frequency band into two input layers.

This choice was motivated by the observed positive correlation between theta power and cognitive workload levels, as well as the existence of a negative correlation for alpha power. Consequently, we segregated these two spectral bands, aimed at scanning the detailed EEG power structure and augmenting the recognition of common spatial-frequency-domain patterns. Fig. 3 provides a graphical representation of the feature arrangement within the input layers of the SCVCNet.

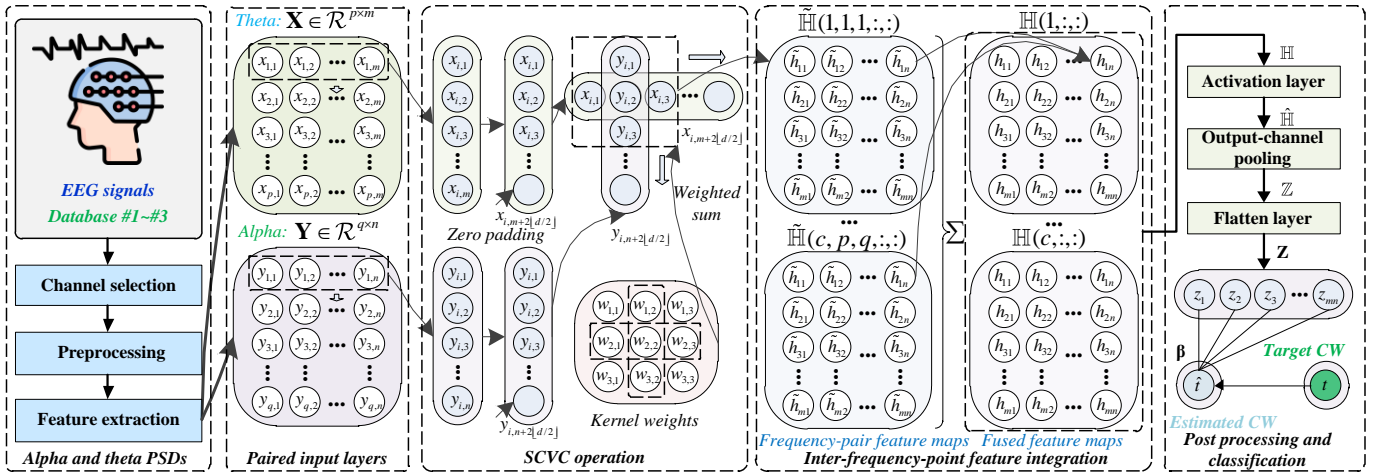

**Fig. 2.** Architecture of the SCVCNet. Cognitive workload is denoted as CW.

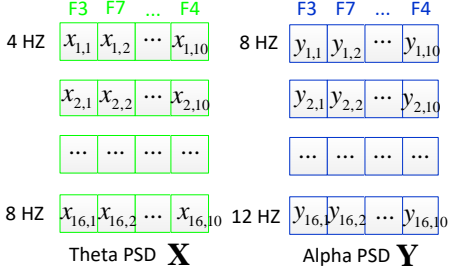

**Fig. 3.** Feature organizations of two input layers of the SCVCNet. The format of the features of the input layer within the theta and alpha bands are shown. The order of the channels along the columns is F3, F7, T7, P7, O1, O2, P8, T8, F8, and F4.

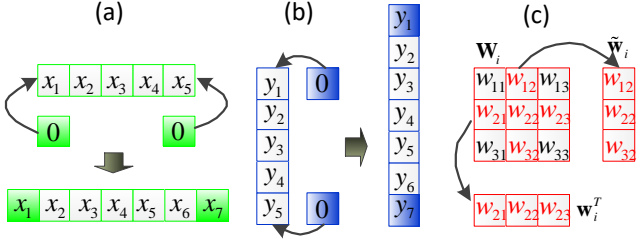

**Fig. 4.** Procedure of the padding and weight copying. In subfigures (a) and (b), we show how to pad the feature vectors from the input layer #1 and #2, respectively. In subfigure (c), we show how to copy the central vectors of the kernel weight.

### B. SCVC Operation

The SCVC operation implements a bilinear transformation through a convolution operation to build EEG feature representations. The SCVC operation encodes the feature between a specified frequency pair. Given the pair of the theta and alpha PSDs of multiple EEG channels defined by $\mathbf{x}_i^T = [x_{i,1}, x_{i,2}, ..., x_{i,m}] \in \mathcal{R}^m$ and $\mathbf{y}_i = [y_{i,1}, y_{i,2}, ..., y_{i,n}]^T \in \mathcal{R}^n$ from a frequency point indexed by $i$, the SCVC can be defined via a feedforward operator,

$$\mathbf{H}_i = SCVC(\mathbf{x}_i^T, \mathbf{y}_i; \mathbf{W}_i, s). \quad (1)$$

In the equation, $\mathbf{H}_i \in \mathcal{R}^{m \times n}$ represents the information from the EEG features $\mathbf{x}_i^T$ and $\mathbf{y}_i$ with a stride of $s$. The matrix $\mathbf{W}_i \in \mathcal{R}^{d \times d}$ is the SCVC kernel weight with the size of $d$. The computation of the operator $SCVC(\cdot, \cdot; \mathbf{W}_i, s)$ is shown in the proposed Algorithm 1.

In lines 1~5, the EEG feature vectors are padded with the zeros at the beginning and the end of each vector. The number of the padded zeros is determined as $\lfloor d/2 \rfloor$, where the kernel size $d$ must be an odd number. The padding operation is illustrated in Fig. 4(a)–(b). In these subfigures, an example is shown with the kernel size $d=3$ and the feature numbers $m = n = 5$. In lines 6~9, we copy the central row $\mathbf{w}_i^T$ and column $\tilde{\mathbf{w}}_i$ from the kernel matrix $\mathbf{W}_i$ with the index of $\lceil d/2 \rceil$. This procedure is illustrated in Fig. 4(c).

In lines 10~18, the computation of the SCVC is performed between the $\mathbf{x}_i^T$, $\mathbf{y}_i$, $\mathbf{w}_i^T$, and $\tilde{\mathbf{w}}_i$. In detail, we first calculate

**Algorithm 1.** Feedforward computation of the SCVC operation.

| $\mathbf{H}_i = SCVC(\mathbf{x}_i^T, \mathbf{y}_i; \mathbf{W}_i, s)$ |
|---|
| **Input:** Theta PSD values $\mathbf{x}_i^T \in \mathcal{R}^m$ |
| Alpha PSD values $\mathbf{y}_i \in \mathcal{R}^n$ |
| Kernel weight $\mathbf{W}_i \in \mathcal{R}^{d \times d}$ |
| Index of the frequency point $i$ |
| Stride $s$ |
| **Output:** Feature representation $\mathbf{H}_i \in \mathcal{R}^{m \times n}$ |

| 1 | Initialize $\mathbf{H}_i \in \mathcal{R}^{m \times n}$ with zeros |
|---|---|
| 2 | Build $\mathbf{z}_i \leftarrow [z_{i,1}, z_{i,2}, ..., z_{i,\lfloor d/2 \rfloor}]$ with zeros |
| 3 | Build $\tilde{\mathbf{z}}_i \leftarrow [\tilde{z}_{i,1}, \tilde{z}_{i,2}, ..., \tilde{z}_{i,\lfloor d/2 \rfloor}]$ with zeros |
| 4 | $\mathbf{x}_i \leftarrow [\mathbf{z}_i, \mathbf{x}_i, \tilde{\mathbf{z}}_i]^T$ |
| 5 | $\mathbf{y}_i \leftarrow [\mathbf{z}_i^T, \mathbf{y}_i, \tilde{\mathbf{z}}_i^T]$ |
| 6 | Extract the central row of $\mathbf{W}_i \in \mathcal{R}^{d \times d}$ |
|  | with the index $\lceil d/2 \rceil$ as $\mathbf{w}_i^T$ |
| 7 | Extract the central column of $\mathbf{W}_i \in \mathcal{R}^{d \times d}$ |
|  | with the index $\lceil d/2 \rceil$ as $\tilde{\mathbf{w}}_i$ |
| 8 | $[w_{\lfloor d/2 \rfloor,1}, w_{\lfloor d/2 \rfloor,2}, ..., w_{\lfloor d/2 \rfloor,d}] \leftarrow \mathbf{w}_i^T$ |
| 9 | $[w_{1,\lceil d/2 \rceil}, w_{2,\lceil d/2 \rceil}, ..., w_{d,\lceil d/2 \rceil}]^T \leftarrow \tilde{\mathbf{w}}_i$ |
| 10 | **for** $k \leftarrow \lfloor d/2 \rfloor + 1 : s : \lfloor d/2 \rfloor + m$ |
| 11 | **for** $l \leftarrow \lfloor d/2 \rfloor + 1 : s : \lfloor d/2 \rfloor + n$ |
| 12 | $h_{kl,row} \leftarrow w_{\lceil d/2 \rceil,1} x_{k-\lfloor d/2 \rfloor} +$ |
|  | $w_{\lceil d/2 \rceil,2} x_{k-\lfloor d/2 \rfloor+1} + ... + w_{\lceil d/2 \rceil,d} x_{k+\lfloor d/2 \rfloor}$ |
| 13 | $h_{kl,col} \leftarrow w_{1,\lceil d/2 \rceil} y_{l-\lfloor d/2 \rfloor} +$ |
|  | $w_{2,\lceil d/2 \rceil} y_{l-\lfloor d/2 \rfloor+1} + ... + w_{d,\lceil d/2 \rceil} y_{l+\lfloor d/2 \rfloor}$ |
| 14 | $h_{kl} \leftarrow h_{kl,row} + h_{kl,col}$ |
| 15 | $h_{kl} \leftarrow h_{kl} + w_{\lceil d/2 \rceil,\lceil d/2 \rceil} x_k y_l -$ |
|  | $w_{\lceil d/2 \rceil,\lceil d/2 \rceil} x_k - w_{\lceil d/2 \rceil,\lceil d/2 \rceil} y_l$ |
| 16 | $\mathbf{H}_i(k, l) \leftarrow h_{kl}$ |
| 17 | **end for** |
| 18 | **end for** |
| 19 | **Return** $\mathbf{H}_i$ |

Note: Theta PSD values $\mathbf{x}_i^T \in \mathcal{R}^m$ are with $m$ EEG channels. Alpha PSD values $\mathbf{y}_i \in \mathcal{R}^n$ are with $n$ EEG channels. Kernel weight $\mathbf{W}_i \in \mathcal{R}^{d \times d}$ is with the size of $d$.

the inner product $h_{kl,row}$ between the central row $\mathbf{w}_i^T$ of the kernel and the elements of the same size in feature vector $\mathbf{x}_i^T$, where the indices of these elements are from $k - \lfloor d/2 \rfloor$ to $k + \lfloor d/2 \rfloor$,

$$h_{kl,row} = w_{\lceil d/2 \rceil,1} x_{k-\lfloor d/2 \rfloor} + w_{\lceil d/2 \rceil,2} x_{k-\lfloor d/2 \rfloor+1} + ... + w_{\lceil d/2 \rceil,d} x_{k+\lfloor d/2 \rfloor}. \quad (2)$$

In the equation, $k$ and $l$ denote the row and column indices of the current entry in the output feature map $\mathbf{H}_i$, respectively. Next, we compute the inner product $h_{kl,col}$ between the central column $\tilde{\mathbf{w}}_i$ of the kernel and the elements of the same size in feature vector $\mathbf{y}_i$, where the elements indices are from $l - \lfloor d/2 \rfloor$ to $l + \lfloor d/2 \rfloor$,

$$h_{kl,col} = w_{1,\lceil d/2 \rceil} y_{l-\lfloor d/2 \rfloor} + w_{2,\lceil d/2 \rceil} y_{l-\lfloor d/2 \rfloor+1} + ... + w_{d,\lceil d/2 \rceil} y_{l+\lfloor d/2 \rfloor}. \quad (3)$$

Then, the sum $h_{kl}$ of the two products is derived as,

$$h_{kl} = h_{kl,row} + h_{kl,col}. \quad (4)$$

Note that the cross term $w_{\lceil d/2 \rceil,\lceil d/2 \rceil} x_k + w_{\lceil d/2 \rceil,\lceil d/2 \rceil} y_l$ in $h_{kl}$ should be updated with $w_{\lceil d/2 \rceil,\lceil d/2 \rceil} x_k y_l$,

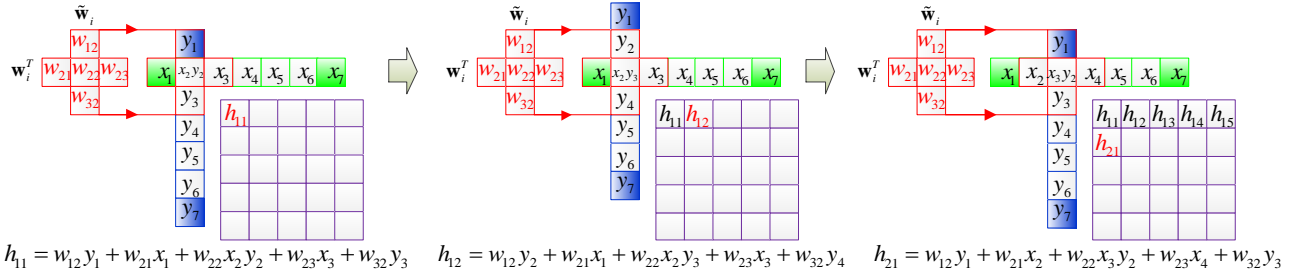

$h_{11} = w_{12}y_1 + w_{21}x_1 + w_{22}x_2x_2 + w_{23}x_3 + w_{32}y_3$ $\quad$ $h_{12} = w_{12}y_2 + w_{21}x_1 + w_{22}x_2 + w_{23}x_3 + w_{32}y_4$ $\quad$ $h_{21} = w_{12}y_1 + w_{21}x_2 + w_{22}x_3y_2 + w_{23}x_4 + w_{32}y_3$

**Fig. 5.** Procedure of the sliding cross-vector convolution operation.

**Algorithm 2.** Feedforward computation of the inter-frequency-point feature integration (IFPFI).

| $\mathbb{H} = IFPFI(\mathbf{X}, \mathbf{Y}; \mathbb{W}, s_m, \mathbf{b})$. | |
|---|---|
| **Input:** | Theta PSD values $\mathbf{X} \in \mathcal{R}^{p \times m}$ |
| | Alpha PSD values $\mathbf{Y} \in \mathcal{R}^{q \times n}$ |
| | Kernel weight $\mathbb{W} \in \mathcal{R}^{c \times p \times q \times d \times d}$ |
| | Stride for the IFPFI $s_m$ |
| | Bias of the IFPFI $\mathbf{b} \in \mathcal{R}^c$ |
| **Output:** | Feature representation $\mathbb{H} \in \mathcal{R}^{c \times m \times n}$ |
| 1 | Initialize $\mathbb{H} \in \mathcal{R}^{c \times m \times n}$ with zeros |
| 2 | Initialize $\tilde{\mathbb{H}} \in \mathcal{R}^{c \times p \times q \times m \times n}$ with zeros |
| 3 | **for** $k = 1:1:c$ |
| 4 |     **for** $i = 1:1:p$ |
| 5 |         **for** $j = 1:1:q$ |
| 6 |             Get $\mathbf{x}_i^T \in \mathcal{R}^m$ |
| 7 |             Get $\mathbf{y}_j \in \mathcal{R}^n$ |
| 8 |             $\mathbf{W}_{kij} \leftarrow \mathbb{W}(k,i,j,:,:)$ |
| 9 |             $\mathbf{H}_{kij} \leftarrow SCVC(\mathbf{x}_i^T, \mathbf{y}_j; \mathbf{W}_{kij}, s_m)$ |
| 10 |             $\tilde{\mathbb{H}}(k,i,j,:,:) \leftarrow \mathbf{H}_{kij}$ |
| 11 |         **end for** |
| 12 |     **end for** |
| 13 |     Select the $k$th entry from $\mathbf{b}$ as $b_k$ |
| 14 |     $\mathbb{H}(k,:,:) \leftarrow \sum_{i=1}^{p}\sum_{j=1}^{q}\tilde{\mathbb{H}}(k,i,j,:,:) + b_k$ |
| 15 | **end for** |
| 16 | **return** $\mathbb{H}$ |

Note: Theta PSD values $\mathbf{X} \in \mathcal{R}^{p \times m}$ are with the length of $p$ from $m$ EEG channels. Alpha PSD values $\mathbf{Y} \in \mathcal{R}^{q \times n}$ with the length of $q$ from $n$ EEG channels. Kernel weight $\mathbb{W} \in \mathcal{R}^{c \times p \times q \times d \times d}$ is with the size of $d$ and output channel $c$ .

$$\tilde{h}_{kl} = h_{kl} + w_{\lceil d/2 \rceil, \lfloor d/2 \rfloor} x_k y_l - w_{\lfloor d/2 \rfloor, \lceil d/2 \rceil} x_k - w_{\lceil d/2 \rceil, \lfloor d/2 \rfloor} y_l . \quad (5)$$

Finally, the SCVC is performed in the 2-D cross-shape with a stride $s$ . In Fig. 5, we show how to compute the SCVC operation with $d = 3$ , $m = n = 5$ , and $s = 1$ as an example.

### C. Inter-frequency-point Feature Integration

To tackle multiple frequency points, the SCVC is applied in EEG frequency pairs and the learned feature maps are fused as shown in Algorithm 2. As described in previous Section, the EEG features from the theta and alpha bands are prepared as $\mathbf{X} \in \mathcal{R}^{p \times m}$ and $\mathbf{Y} \in \mathcal{R}^{q \times n}$ , respectively. This inter-frequency-point feature integration (IFPFI) module could derive the representation of $c$ output channels to improve the richness of the learned frequency structures. The details of the Algorithm 2 are described as follows.

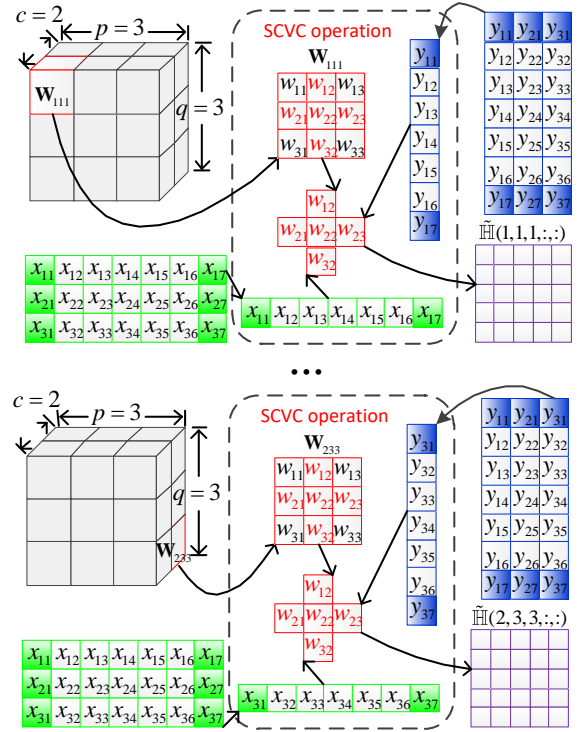

**Fig. 6.** Procedure for computing the frequency-pair feature representations in the IFPFI module.

In lines 1~2, we initialize two empty tensors to store the fused feature maps $\mathbb{H} \in \mathcal{R}^{c \times m \times n}$ and that from frequency pairs $\tilde{\mathbb{H}} \in \mathcal{R}^{c \times p \times q \times m \times n}$ . In the two inner loops of lines 4~12, the SCVC operation in Algorithm 1 is carried out between the $i$th EEG frequency point in the theta feature matrix $\mathbf{X}$ and the $j$th EEG frequency point in the alpha feature matrix $\mathbf{Y}$ . The selected feature vectors are represented as $\mathbf{x}_i^T \in \mathcal{R}^m$ and $\mathbf{y}_j \in \mathcal{R}^n$ . In line 8, the weight matrix $\mathbf{W}_{kij} \in \mathcal{R}^{d \times d}$ is taken out from the kernel weight tensor $\mathbb{W} \in \mathcal{R}^{c \times p \times q \times d \times d}$ , where $k$ is the index of the output channel. In line 9, the SCVC is performed based on $\mathbf{x}_i^T$ , $\mathbf{y}_j$ , $\mathbf{W}_{kij}$ , and the predefined stride $s_m$ ,

$$\mathbf{H}_{kij} = SCVC(\mathbf{x}_i^T, \mathbf{y}_j; \mathbf{W}_{kij}, s_m). \quad (6)$$

In the equation, $\mathbf{H}_{kij} \in \mathcal{R}^{m \times n}$ denotes the learned feature map of the $k$th output channel between the $i$th and $j$th EEG frequency point in theta and alpha bands, respectively. The derived $\mathbf{H}_{kij}$ is then restored in $\tilde{\mathbb{H}}$ . In the outer loop, we sum the $\tilde{\mathbb{H}}$ entries along the dimensions of the EEG frequency domain,

$$\mathbb{H}(k,:,:) = \sum_{i=1}^{p}\sum_{j=1}^{q}\tilde{\mathbb{H}}(k,i,j,:,:) + b_k, \quad (7)$$

where $b_k$ is a scalar bias. In the end, the summed feature map is returned as $\mathbb{H} \in \mathcal{R}^{c \times m \times n}$ for all $c$ output channels. In Fig. 6, we show the procedure for computing frequency pair feature representations in the IFPFI with $c = 2$, $m = n = 5$, $p = q = 3$, $d = 3$, and $s_m = 1$ as an example.

### D. Post Processing and Classification Layers

Given $N$ distinct input samples $(\mathbb{X}, \mathbb{Y})$ of the EEG features $\mathbb{X} \in \mathcal{R}^{p \times m \times N}$ and $\mathbb{Y} \in \mathcal{R}^{q \times n \times N}$ derived from the theta and alpha bands, we initialize the kernel weight $\mathbb{W}^r$ and the bias $\mathbf{b}^r$ in the SCVC and IFPFI with random values. The IFPFI computes the output $\mathbb{H}^r \in \mathcal{R}^{c \times m \times n \times N}$ according to Algorithm 2 as follows,

$$\mathbb{H}^r = IFPFI(\mathbb{X}, \mathbb{Y}; \mathbb{W}^r, s_m, \mathbf{b}^r). \tag{8}$$

The kernel weight and the bias are drawn from the modified uniform distribution as indicated in [28], respectively.

Next, we apply the logistic sigmoid activation function $G(\mathbb{H}^r)$ on $\mathbb{H}^r$ to derive the neuron activations $\hat{\mathbb{H}}^r \in \mathcal{R}^{c \times m \times n \times N}$,

$$\hat{\mathbb{H}}^r = G(\mathbb{H}^r) = 1 / (1 + e^{\mathbb{H}^r}). \tag{9}$$

Considering the dimension of the tensor $\hat{\mathbb{H}}^r$ can be very large, we used a output-channel pooling layer $CP(\cdot)$ to reduce the dimensionality,

$$\mathbb{Z} = CP(\hat{\mathbb{H}}^r) = 1 / c \sum_{j=1}^{c} \hat{\mathbb{H}}^r(j, :, :, :). \tag{10}$$

Here, $\mathbb{Z} \in \mathcal{R}^{m \times n \times N}$ denotes the reduced feature representations of $N$ EEG samples. Then, a flatten layer $\mathbf{Z} = FL(\mathbb{Z})$ is employed to reshape the tensor $\mathbb{Z}$ as a matrix $\mathbf{Z} \in \mathcal{R}^{(mn) \times N}$.

Finally, we employed a ridge regression layer as the cognitive workload classifier to assigned the predicted workload labels $\hat{\mathbf{t}} \in \mathcal{R}^N$ for the flattened feature representations $\mathbf{Z}$,

$$\hat{\mathbf{t}} = \mathbf{Z}\boldsymbol{\beta}, \tag{11}$$

where $\boldsymbol{\beta} \in \mathcal{R}^{(mn)}$ is the regression weight that required to be tuned. Each entry $\hat{t}$ in the vector $\hat{\mathbf{t}}$ could be classified by a threshold 0.5 as 1 (loaded class) and 0 (unloaded class).

**Algorithm 3.** Training the SCVCNet.

| | $\boldsymbol{\beta} = SCVCNet\_train\left((\mathbb{X}, \mathbb{Y}, \mathbf{t}), s_m, C\right)$ |
|---|---|
| **Input:** | $N$ samples $\mathbb{X} \in \mathcal{R}^{p \times m \times N}$ |
| | $N$ samples $\mathbb{Y} \in \mathcal{R}^{q \times n \times N}$ |
| | $N$ target labels $\mathbf{t} \in \mathcal{R}^N$ |
| | Stride of the IFPFI $s_m$ |
| | Hyperparameter $C$ |
| **Output:** | Ridge regression weight $\boldsymbol{\beta} \in \mathcal{R}^{(mn)}$ |
| 1 | Randomly assign kernel weight $\mathbb{W}^r$ |
| 2 | Randomly assign bias $\mathbf{b}^r$ |
| 3 | $\mathbb{H}^r \leftarrow IFPFI(\mathbb{X}, \mathbb{Y}; \mathbb{W}^r, s_m, \mathbf{b}^r)$ |
| 4 | $\hat{\mathbb{H}}^r \leftarrow G(\mathbb{H}^r)$ |
| 5 | $\mathbb{Z} \leftarrow 1 / c \sum_{j=1}^{c} \hat{\mathbb{H}}^r(j, :, :, :)$ |
| 6 | $\mathbf{Z} \leftarrow FL(\mathbb{Z})$ |
| 7 | **if** $N \le mn$ |
| 8 | $\boldsymbol{\beta} \leftarrow \mathbf{Z}(\mathbf{I} / C + \mathbf{Z}^T \mathbf{Z})^{-1} \mathbf{t}$ |
| 9 | **else if** $N > mn$ |
| 10 | $\boldsymbol{\beta} \leftarrow (\tilde{\mathbf{I}} / C + \mathbf{Z}\mathbf{Z}^T)^{-1} \mathbf{Z}\mathbf{t}$ |
| 11 | **end if** |
| 12 | **return** $\boldsymbol{\beta}$ |

Note: $\mathbb{X} \in \mathcal{R}^{p \times m \times N}$ of theta band is with the length of $p$ from $m$ EEG channels. $\mathbb{Y} \in \mathcal{R}^{q \times n \times N}$ of alpha band is with the length of $q$ from $n$ EEG channels.

### E. Training the SCVCNet

The training of the SCVCNet leverages the hierarchical ELM [29]-[31] algorithm, where the regression weight $\boldsymbol{\beta}$ for mapping the feature representations incorporated with random weights can be directly solved by the regularized least-squares solution,

$$\min_{\boldsymbol{\beta}} \Psi(\boldsymbol{\beta}) = \min_{\boldsymbol{\beta}} (\|\mathbf{Z}\boldsymbol{\beta} - \mathbf{t}\|_2^2 + C\|\boldsymbol{\beta}\|_2^2). \tag{12}$$

The first term denotes the least-square error between the predicted workload labels $\hat{\mathbf{t}} = \mathbf{Z}\boldsymbol{\beta}$ and the target labels $\mathbf{t} \in \mathcal{R}^N$ of the $N$ samples of the EEG features $(\mathbb{X}, \mathbb{Y})$. The second term is used to reduce the model complexity by minimizing the L2 norm of the regression weight $\boldsymbol{\beta}$. That is, some entries in $\boldsymbol{\beta}$ can be closed to zeros when the minimum of $\Psi(\boldsymbol{\beta})$ is achieved and the redundant feature representations can be eliminated. The $C$ is a hyperparameter to balance the two terms.

According to [29], [32]-[34], the $\boldsymbol{\beta}$ can be computed by the closed-form solution,

$$\boldsymbol{\beta} = \mathbf{Z}(\mathbf{I} / C + \mathbf{Z}^T \mathbf{Z})^{-1} \mathbf{t}, N \le mn, \tag{13}$$

$$\boldsymbol{\beta} = (\tilde{\mathbf{I}} / C + \mathbf{Z}\mathbf{Z}^T)^{-1} \mathbf{Z}\mathbf{t}, N > mn. \tag{14}$$

In the equations, $\mathbf{I} \in \mathcal{R}^{N \times N}$ and $\tilde{\mathbf{I}} \in \mathcal{R}^{(mn) \times (mn)}$ are the identity matrices. In Eqn. (13), the inverse matrix exists when the input $N$ samples in $(\mathbb{X}, \mathbb{Y})$ are distinct. In Eqn. (14), the inverse matrix exists when the input $mn$ features in $(\mathbb{X}, \mathbb{Y})$ are distinct.

## V. RESULTS

### A. Validation Approaches

To validate the generalization capability of the cross-task and inter-individual set cognitive workload recognizers based on the N-back, STEW, and EEGMAT databases, we employed the following methodology: from each phase (resting or task) across all participants within a given database, we selectively extracted the initial 20% of feature vectors in accordance with their temporal sequencing. These extracted vectors collectively constitute three distinct validation sets, namely N-back-VA, STEW-VA, and EEGMAT-VA. Conversely, the remaining 80% of feature vectors are designated as N-back\VA, STEW\VA, and EEGMAT\VA. Subsequently, we explored two training-testing paradigms, encompassing six unique cases.

In Paradigm 1, we employ the validation set (-VA) derived from a specific database for the purpose of fine-tuning the classifier's hyperparameters. Subsequently, the remaining feature vectors (\VA) from the same database are designated as the testing set, while an alternative database is chosen to serve as the training set. In contrast, Paradigm 2 involves the utilization of the entirety of a given database's features as the

TABLE II
TRAINING AND TESTING CASES UNDER TWO PARADIGMS.

| | Training set | Validation set | Testing Set #1 | Testing Set #2 |
|---|---|---|---|---|
| Case 1 | N-back | STEW-VA | STEW\VA | EEGMAT |
| Case 2 | N-back | EEGMAT-VA | EEGMAT\VA | STEW |
| Case 3 | STEW | N-back-VA | N-back\VA | EEGMAT |
| Case 4 | STEW | EEGMAT-VA | EEGMAT\VA | N-back |
| Case 5 | EEGMAT | N-back-VA | N-back\VA | STEW |
| Case 6 | EEGMAT | STEW-VA | STEW\VA | N-back |

Note: Testing Sets #1 and #2 denote the testing sets defined under Paradigms 1 and 2, respectively.

TABLE III
HYPER-PARAMETER SETTINGS OF THE SCVCNET.

| Hyper-parameters | Case 1 | Case 2 | Case 3 | Case 4 | Case 5 | Case 6 |
|---|---|---|---|---|---|---|
| Random seed | 42 | 42 | 42 | 42 | 42 | 42 |
| Output channels $c$ | 76 | 77 | 6 | 86 | 77 | 88 |
| Kernel size $d$ | 3 | 3 | 3 | 3 | 3 | 3 |
| Conv groups | 1 | 1 | 1 | 1 | 1 | 1 |
| Bias | With | With | With | With | With | With |
| Activation | Sigm | Sigm | Sigm | Sigm | Sigm | Sigm |
| Initial weight | Uni | Uni | Uni | Uni | Uni | Uni |
| Regularization C | $1.5 \times 10^{-10}$ | $3.1 \times 10^{-10}$ | $8.5 \times 10^{-9}$ | $1.0 \times 10^{-4}$ | $6.2 \times 10^{-7}$ | $3.38 \times 10^{7}$ |

Note: The abbreviations Sigm and Uni denote the logistic sigmoid function and uniform distribution, respectively.

TABLE IV
PARTICIPANT-AVERAGE ACCURACY AND F1 SCORE ON THE TESTING
SET OF THE SCVCNET.

| | Paradigm 1 | | Paradigm 2 | |
|---|---|---|---|---|
| | ACC | F1 | ACC | F1 |
| Case 1 | 0.6797 | 0.6752 | 0.6389 | 0.6220 |
| Case 2 | 0.6528 | 0.6437 | **0.6854** | **0.6784** |
| Case 3 | 0.6875 | 0.6875 | 0.5639 | 0.5510 |
| Case 4 | 0.6736 | 0.6658 | 0.5900 | 0.5729 |
| Case 5 | 0.6938 | 0.6833 | 0.6292 | 0.5914 |
| Case 6 | **0.7005** | **0.6905** | 0.6300 | 0.6300 |
| Mean | 0.6813 ±0.0155 | 0.6743 ±0.0160 | 0.6229 ±0.0383 | 0.6076 ±0.0416 |

Note: The highest value in each column is marked in bold.

testing set, with the validation and training sets independently sourced from the remaining two databases. The intricate details of these six distinct cases are comprehensively presented in Table II. Ultimately, we gauge the model's performance using participant-averaged metrics, specifically the accuracy (ACC) and macro-F1 score (F1), computed on the designated testing sets.

### B. Algorithm Implementation and Hyperparameters

All models were trained using an NVIDIA GTX 1650 GPU, equipped with CUDA 10.2, and implemented in Python 3.9.7, utilizing either PyTorch 2.0.0 or TensorFlow 2.10.0. Access to the EEG features from the three databases and the associated codebase is available through our website: https://github.com/7ohnKeats/SCVCNet. Table III presents a comprehensive overview of the hyperparameter configurations for the SCVCNet. In the interest of result reproducibility, we adhered to a fixed random seed strategy for all cases. These parameters encompassed crucial aspects such as determining the optimal number of output channels, kernel size, the number of convolutional groups, and regularization hyper-parameters.

### C. Performance of the SCVCNet

In Table IV, we show the performance of the SCVCNet. Overall, the participant-average values of the ACC are higher than the F1 for all cases and paradigms. The case-average ACC under Paradigm 1 (0.6813) is higher than that of Paradigm 2 (0.6229). The reason behind this is that validation and testing data under Paradigm 1 were selected from the same database. The selection of the hyperparameters for the SCVCNet inclines to better fit the testing feature distribution. Under Paradigm 1, the SCVCNet under Case 6 achieves the highest ACC (0.7005) and F1 (0.6905) among all cases, where the training, validation, and testing sets are EEGMAT, STEW-VA, and STEW\VA, respectively. Under Paradigm 2, the SCVCNet under Case 2 elicits the highest ACC (0.6854) and F1 (0.6784) in all cases, where the training, validation, and testing sets are N-back, EEGMAT-VA, and STEW, respectively.

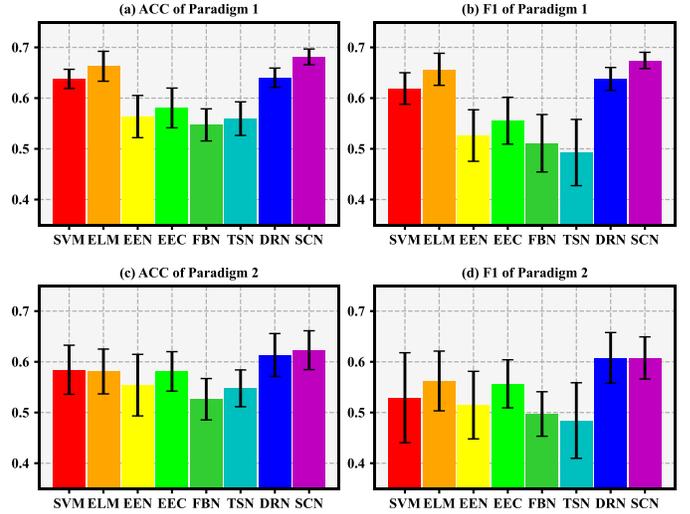

**Fig. 7.** Case-average testing performance of different cognitive workload classifiers. The abbreviations EEN, EEC, FBN, TSN, DRN, and SCN, denote EEGNet, EEG Conformer, FBMSNet, TS-SEFFNet, DRNANet, and SCVCNet, respectively.

### D. Performance Comparison

We then compare the performance of the SCVCNet with several classical classifiers and state-of-the-art deep neural networks used for EEG cognitive pattern recognition or brain-computer interface, including SVM, ELM [30], EEGNet [35], EEG Conformer [36], FBMSNet [37], TS-SEFFNet [38], and DRNANet [15]. The case-average performance of the classifiers examined by each paradigm is illustrated in Fig. 7. Overall, the SCVCNet achieves the highest ACC (0.6813) and F1 (0.6743) under Paradigms 1. Under paradigm 2, the SCVCNet's ACC (0.6229) is superior to all other classifiers while the DRNANet achieves the highest F1 (0.6077) and the SCVCNet shows the second highest value (0.6076). For other classifiers, ELM, SVM, and DRNA-Net show higher performance than that of the EEGNet, EEG Conformer, FBMSNet, and TS-SEFFNet under Paradigm 1. It implies deeper network structures may induce potential overfitting on the training set and reduce the classifier's generalizability considering the domain adaptation techniques were not implemented.

In Tables V-VII, we separately list the participant-average performance of all classifiers of two paradigms in each case. Under paradigm 1, the SCVCNet shows the highest ACC and F1 in Cases 3, 4, and 5. Under paradigm 2, the SCVCNet elicits the highest performance in Cases 2 and 6. In the remaining cases, the SCVCNet could be comparable to the optimal classifier. Note that other deep neural networks may show low robustness across different cases. For instance, the EEG Conformer achieves an ACC 0.6285 under Case 2 in Table V. However, this value (0.5188) is close to the random guess (0.5000) under Case 3 in Table VI. The DRNANet shows stable performance under Paradigm 2 for all cases while the performance decreases in Cases 3 and 6 under Paradigm 1.

## TABLE V
### PARTICIPANT-AVERAGE ACCURACY AND F1 SCORE ACROSS DIFFERENT COGNITIVE WORKLOAD CLASSIFIERS EXAMINED BY CASES 1 AND 2.

| Classifier | Paradigm 1 ACC | F1 | Paradigm 2 ACC | F1 | Paradigm 1 ACC | F1 | Paradigm 2 ACC | F1 |
|---|---|---|---|---|---|---|---|---|
| SVM | 0.6432 | 0.6200 | 0.6444 | 0.6443 | **0.6632** | **0.6629** | 0.6542 | 0.6327 |
| ELM | **0.6927** | **0.6874** | 0.6167 | 0.5998 | 0.6597 | 0.6596 | 0.6500 | 0.6462 |
| EEGNet | 0.5026 | 0.5022 | 0.5444 | 0.4712 | 0.5660 | 0.4965 | 0.4375 | 0.4170 |
| EEGConf | 0.5755 | 0.5215 | 0.6278 | 0.6226 | 0.6285 | 0.6229 | 0.5771 | 0.5235 |
| FBMSNet | 0.5781 | 0.5765 | 0.5250 | 0.5069 | 0.5417 | 0.4565 | 0.5167 | 0.4888 |
| SEFFNet | 0.4974 | 0.3996 | 0.5972 | 0.5800 | 0.6042 | 0.5849 | 0.4979 | 0.4011 |
| DRANet | 0.6563 | 0.6561 | **0.6472** | **0.6449** | 0.6424 | 0.6404 | 0.6667 | 0.6666 |
| SCVC | 0.6797 | 0.6752 | 0.6389 | 0.6220 | 0.6528 | 0.6437 | **0.6854** | **0.6784** |

Note: EEGConf and SEFFNet denote the EEG Conformer and the TS-SEFFNet, respectively. The left and right four columns are the performance under Case 1 and 2, respectively.

## TABLE VI
### PARTICIPANT-AVERAGE ACCURACY AND F1 SCORE ACROSS DIFFERENT COGNITIVE WORKLOAD CLASSIFIERS EXAMINED BY CASES 3 AND 4.

| Classifier | Paradigm 1 ACC | F1 | Paradigm 2 ACC | F1 | Paradigm 1 ACC | F1 | Paradigm 2 ACC | F1 |
|---|---|---|---|---|---|---|---|---|
| SVM | 0.6438 | 0.6131 | 0.5472 | 0.4997 | 0.6389 | 0.6388 | 0.5250 | 0.3866 |
| ELM | 0.6563 | 0.6446 | 0.5306 | 0.4589 | 0.6076 | 0.5959 | 0.5250 | 0.5344 |
| EEGNet | 0.5250 | 0.4473 | 0.6139 | 0.6117 | 0.6007 | 0.5981 | 0.5450 | 0.4751 |
| EEGConf | 0.5188 | 0.5172 | 0.6250 | 0.6185 | 0.6181 | 0.6105 | 0.5250 | 0.5240 |
| FBMSNet | 0.4813 | 0.4226 | 0.5389 | 0.5003 | 0.5521 | 0.5428 | 0.4450 | 0.4154 |
| SEFFNet | 0.5625 | 0.4852 | 0.5806 | 0.5698 | 0.5764 | 0.5663 | 0.5650 | 0.4938 |
| DRANet | 0.5892 | 0.5892 | **0.6278** | **0.6277** | 0.6458 | 0.6457 | **0.6100** | **0.6033** |
| SCVC | **0.6875** | **0.6875** | 0.5639 | 0.5510 | **0.6736** | **0.6658** | 0.5900 | 0.5729 |

Note: The left and right four columns are the performance under Case 3 and 4, respectively.

## TABLE VII
### PARTICIPANT-AVERAGE ACCURACY AND F1 SCORE ACROSS DIFFERENT COGNITIVE WORKLOAD CLASSIFIERS EXAMINED BY CASES 5 AND 6.

| Classifier | Paradigm 1 ACC | F1 | Paradigm 2 ACC | F1 | Paradigm 1 ACC | F1 | Paradigm 2 ACC | F1 |
|---|---|---|---|---|---|---|---|---|
| SVM | 0.6000 | 0.5604 | 0.5750 | 0.4835 | 0.6380 | 0.6184 | 0.5600 | 0.5281 |
| ELM | 0.6625 | 0.6624 | 0.5833 | 0.5674 | 0.6979 | **0.6909** | 0.5800 | 0.5766 |
| EEGNet | 0.5625 | 0.5363 | 0.6229 | 0.5751 | 0.6250 | 0.5765 | 0.5600 | 0.5376 |
| EEGConf | 0.5438 | 0.5046 | 0.5917 | 0.5500 | 0.5990 | 0.5556 | 0.5400 | 0.5009 |
| FBMSNet | 0.5625 | 0.5014 | 0.5667 | 0.5652 | 0.5677 | 0.5663 | 0.5650 | 0.5062 |
| SEFFNet | 0.5438 | 0.4395 | 0.5354 | 0.4749 | 0.5729 | 0.4804 | 0.5100 | 0.3356 |
| DRANet | 0.6563 | 0.6552 | 0.5938 | **0.5934** | 0.6406 | 0.6404 | 0.5350 | 0.5102 |
| SCVC | **0.6938** | **0.6833** | **0.6292** | 0.5914 | **0.7005** | 0.6905 | **0.6300** | **0.6300** |

Note: The left and right four columns are the performance under Case 5 and 6, respectively.

### E. Ablation Study

To verify the validity of the proposed SCVC operation and the random-weights training method, we built two reduced models of the SCVCNet and compared the accuracies with the full version. Figure 8 illustrates the participant-average accuracy of the three models, i.e., the full version (FU), reduced version #1 (R1), and reduced version #2 (R2), under six cases and two paradigms (denoted and t1 and t2). In the R1 model, the SCVC feedforward architecture in the SCVCNet was replaced with a fully connected layer. In the R2 model, the training algorithm with the random weights in the SCVCNet was replaced with the ADAM approach with 100 training epochs. To facilitate a fair comparison, the remaining architectures and hyperparameters of the three models are the same. Overall, the case-average accuracies satisfy FUt1 (0.6813) > R2t1 (0.6512) > R1t1 (0.6201) and FUt2 (0.6229) > R2t2 (0.5918) > R1t2 (0.5776), which is consistent with most cases shown in the figure. It implies the effectiveness of implementing both Algorithms 2 and 3 in the SCVCNet.

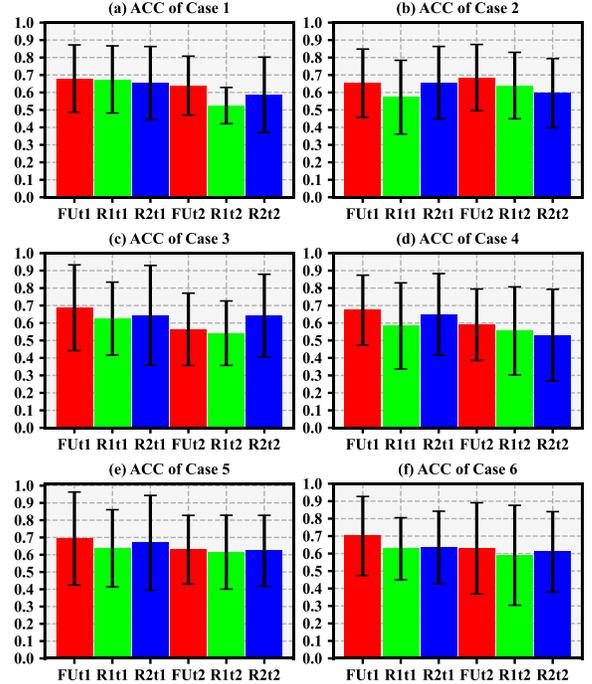

**Fig. 8.** Participant-average accuracy of the full version (FU), reduced version #1 (R1), and reduced version #2 (R2) of the SCVCNet, under six cases (a)–(f) and two paradigms (denoted and t1 and t2).

### F. SCVCNet's Performance on Different Random Seeds

To investigate the impact of randomly selecting SCVC weight configurations on its performance, we conducted twenty rounds of performance metric calculations for SCVCNet. Each round involved a distinct seed (the seed's values are from 1 to 20) to initialize the model's weights. Subsequently, the average performance metrics are presented in Table VIII. The standard deviations of accuracy across all rounds were approximately 1.8% and 3.3% for Paradigms 1 and 2, respectively. Regarding the F1 score, the corresponding standard deviations were 2.0% and 4.0%. We employed a two-tailed Wilcoxon signed-rank test to assess the significance of differences in accuracy and F1 score between Tables IV and VIII. The results indicate that there are

## TABLE VIII
### AVERAGE TESTING ACCURACY AND F1 SCORE OF THE SCVCNET OF TWENTY DIFFERENT RAMDOM SEEDS.

| | Paradigm 1 ACC | F1 | Paradigm 2 ACC | F1 |
|---|---|---|---|---|
| Case1 | **0.7064** ±0.0097 | **0.7008** ±0.0117 | 0.6107 ±0.0186 | 0.6029 ±0.0236 |
| Case2 | 0.6595 ±0.0173 | 0.6562 ±0.0190 | **0.6520** ±0.0275 | **0.6445** ±0.0284 |
| Case3 | 0.6863 ±0.0281 | 0.6806 ±0.0292 | 0.5831 ±0.0334 | 0.5540 ±0.0501 |
| Case4 | 0.6370 ±0.0186 | 0.6256 ±0.0225 | 0.6025 ±0.0534 | 0.5946 ±0.0533 |
| Case5 | 0.6788 ±0.0176 | 0.6719 ±0.0161 | 0.6115 ±0.0239 | 0.5738 ±0.0346 |
| Case6 | 0.6780 ±0.0180 | 0.6676 ±0.0239 | 0.5935 ±0.0436 | 0.5854 ±0.0471 |
| Mean | 0.6743 ±0.0179 | 0.6671 ±0.0204 | 0.6089 ±0.0334 | 0.5925 ±0.0395 |

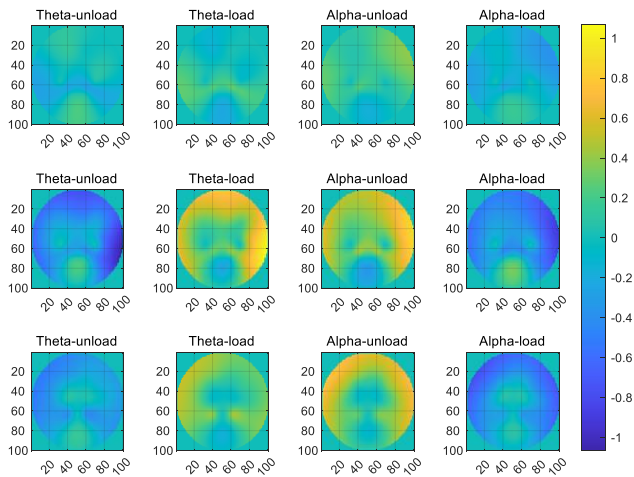

**Fig. 9.** Participant-average PSD feature maps generated by 10 common channels across three databases. For each subfigure, the frequency band and the cognitive workload levels are labeled at the top. The first, second, third rows of the subfigures denote the averaged feature maps from the N-back, STEW, and EEGMAT databases, respectively.

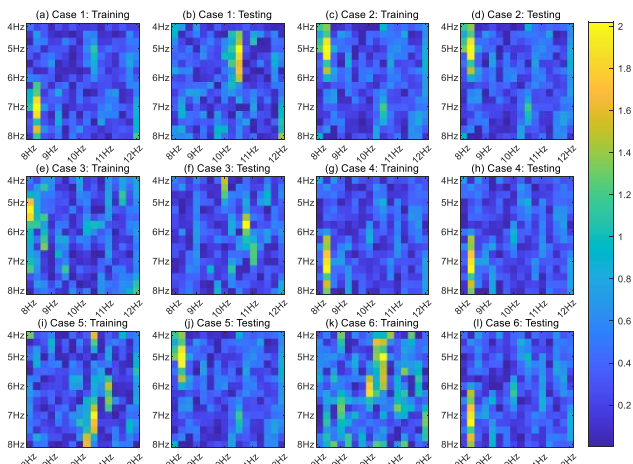

**Fig. 10.** Averaged absolute value of the PSD representation in frequency pairs in theta and alpha bands derived by the SCVC operation. The training set and Testing set #2 in all cases are separately shown.

no statistically significant differences in performance. For accuracy, we obtained *p*-values of 0.56 (Signrank = 14) and 0.22 (Signrank = 17) with Paradigm 1 and 2, respectively. Regarding the F1 score, the derived *p*-values were 0.69 (Signrank = 13) and 0.31 (Signrank = 16) for Paradigm 1 and 2, respectively. These findings suggest that the initialization of random weights does not significantly affect the performance of SCVCNet.

## VI. DISCUSSION

### A. Spatial Common Patterns of the Selected Channels

In Fig. 9, we have depicted the projection of the feature vectors extracted from the common ten channels across three distinct databases onto 2D spatial maps. This projection was achieved by interpolating 100×100 spatial locations, adhering to the 10-20 system within a square configuration. Subsequently, we delineated a circular boundary within the square to represent the scalp's spatial limits, removing any interpolated values that extended beyond this boundary. Within each subfigure, the spatial map corresponds to a specific frequency band, cognitive workload level, and database, with an averaging process applied across all participants. Across all subfigures, a discernible trend emerges: theta power exhibits an increase in the frontal, temporal, and parietal regions as cognitive workload escalates from unloaded to loaded levels, while the opposite pattern manifests itself for alpha power. Despite the relatively diminished discriminative capacity of the spatial maps within the N-back database, it is noteworthy that the spatial distribution of power remains consistent across the N-back, STEW, and EEGMAT databases. This consistency aligns with prior research findings [5], [12], [15], [24], affirming the practicality of cross-task cognitive workload recognition. These works reported that EEG theta oscillations during cognitive processing originate from the dynamic functional binding of widely distributed cortical assemblies [11]. Typically, as task demands increase, theta power exhibits an increase in the midline frontal and parietal regions, while alpha power decreases in the frontal and parietal occipital regions [9].

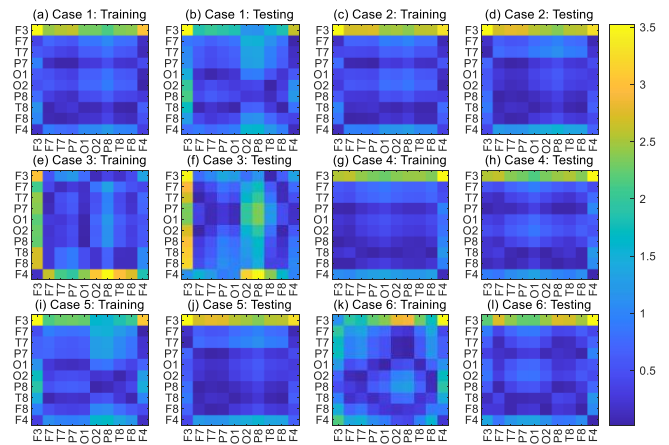

**Fig. 11.** Averaged absolute value of the PSD representation of EEG channel pairs derived by the SCVC operation. The training set and Testing set #2 in all cases are separately shown.

### B. Feature Representation Distribution

In Fig. 10, we present a visual depiction of the absolute values of the averaged PSDs across various frequency points within the theta and alpha bands, as obtained through the SCVC operation. We independently visualize both the training and testing sets for all cases under Paradigm 2. We deliberately omit the results of Paradigm 1 due to its inherent simplicity, given that the validation and testing sets originate from the same database. Notably, we observed strikingly similar patterns between the training and testing sets in Cases 2 and 4, where the low-frequency theta components exhibit significant interaction with the low-frequency alpha components. In Case 1, it remains consistent that features located at 10.5 Hz and 12 Hz within the alpha band demonstrate substantial interaction with both high and low-frequency components within the theta band, respectively. This observation partially underscores the

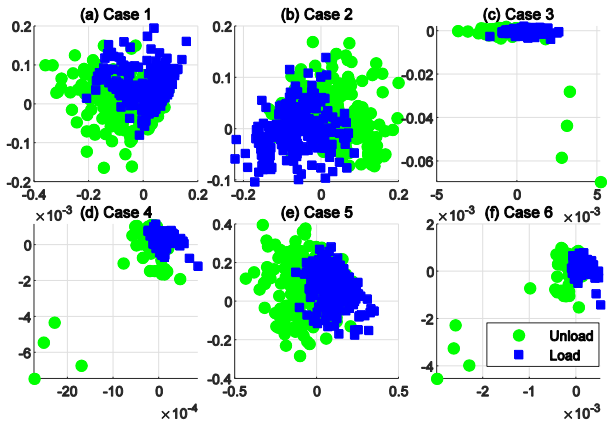

**Fig. 12.** Scatterplots of the 2D embedding mapped by the locality preservation projection from Testing Set #2. Subfigure (a)–(f) denote the results under Cases 1–6, respectively.

TABLE IX
QUALITATIVE COMPARISON WITH REPORTED STUDIES.

| Reference | Classifier | Tasks | IS | Classes | Accuracy |
|---|---|---|---|---|---|
| [18] | MLP | Three | Same | Binary | 44.8% |
| [19] | RF | Two | Same | Binary | 55.6% |
| [21] | SVM | Four | Same | 3-class | 80.3% |
| [22] | SVM | Two | Same | Binary | 87.0% |
| [23] | DBLSTM | Two | Same | Binary | 92.8% |
| [24] | 3DCNetB | Two | Same | Binary | 88.9% |
| [15] | DRNANet | Two | Different | Binary | 60.6% |
| Ours | SCVCNet | Three | Different | Binary | 68.1/62.3% |

Note: The IS denotes whether the individual sets between the training and testing sets are the same or different. The DBLSTM and 3DCNetB denote the deep bidirectional long-short term memory network (BLSTM) and 3D convolutional neural network combined with BLSTM. .

consistent patterns in finer-grained frequency samples. Furthermore, in Fig. 11, we illustrate the average SCVC representations of PSDs originating from the two frequency bands across channel pairs. Channel F3 consistently exhibits heightened values when subjected to the SCVC operation in conjunction with other channels across all cases. This suggests that the fused representations of finer-grained frequency features may exhibit robust patterns to a greater extent, particularly in the left-frontal scalp regions.

### C. Discriminating Capability of the Feature Representations

To assess the discriminative potential of the SCVCNet, which was trained on the EEG PSDs as depicted in Fig. 12, we present a visualization of the 2D embedding generated using the Locality Preserving Projection (LPP) [39] technique based on the acquired feature representations. In Fig. 12, we display the scatterplot of the LPP-mapped embeddings obtained from the testing set under Paradigm 2 for each experimental case. With the exception of Case 2, we observe that samples from the loaded and unloaded classes are predominantly situated in the right and left regions of the 2D feature space, respectively. This observation suggests a certain degree of stability in the SCVCNet's ability to generalize the modeled EEG feature distribution from one cognitive task to another, successfully adapting to different task demands. Nonetheless, inter-class overlap persists, as corroborated by the reported accuracies (averaged with 62.3%) in Section V.

### D. Comparison with Reported Studies

In Table IX, we have compiled a list of prior works centered around the utilization of diverse EEG data sources for cross-task cognitive workload recognition. It is noteworthy that the experimental designs for training and testing within the literature differ from those adopted in the current investigation. Consequently, undertaking a quantitative comparison of the reported accuracies proves unfeasible. In qualitative terms, it is discernible that our proposed methodology surpasses the performance of [15], [18], and [19], albeit falling short in comparison to the methodologies described in [21]-[24]. This discrepancy in performance arises from the fact that our current study employs distinct sets of EEG sources from different tasks for each individual, whereas the aforementioned studies [21]-[24] employ consistent individual sets. This variance in EEG source selection diminishes the influence of inter-individual variability, consequently impacting the classifier's performance. To encapsulate, it is imperative to acknowledge that the assessment of cognitive workload across tasks and among individuals remains a formidable challenge.

### VII. CONCLUSION

In this paper, we proposed SCVCNet, a method based on sliding cross-vector convolution (SCVC) operation, for recognizing cognitive workload levels under cross-task and inter-individual-set conditions. The SCVC operation and IFPFI module effectively integrate finer-grained power information between adjacent frequency samples in alpha and theta PSD feature vectors. By utilizing the N-back, STEW, and EEGMAT databases, we observed similar spatial patterns in the common ten EEG channels between unloaded and loaded levels of cognitive workload. Based on this observation, we conducted extensive experiments, and SCVCNet demonstrated superior performance compared to seven competitive EEG-based workload recognizers. This indicates that the task- and individual-set-related patterns were partially removed.

However, there are still limitations and areas for future work that should be considered. Firstly, the interference from task and individual sets in PSD features has not been completely eliminated. Future works may investigate how to improve the feature filtering capability of the SCVCNet. Secondly, the current databases are unable to distinguish between task-related and individual-set-related EEG sources since these factors vary simultaneously. Future work could involve a decoupling module to quantitatively analyze these two sources of interference.